\documentclass[journal]{vgtc}                

\ifpdf
  \pdfoutput=1\relax                   
  \usepackage{graphicx}                
  \DeclareGraphicsExtensions{.pdf,.png,.jpg,.jpeg} 
\else
  \usepackage{graphicx}                
  \DeclareGraphicsExtensions{.eps}     
\fi%

\graphicspath{{figs/}{pictures/}{images/}{./}} 

\usepackage{microtype}                 
\PassOptionsToPackage{warn}{textcomp}  
\usepackage{textcomp}                  
\usepackage{mathptmx}                  
\usepackage{times}                     
\usepackage{cite}                      
\usepackage{tabu}                      
\usepackage{booktabs}                  

\onlineid{1015}

\vgtccategory{Research}
\vgtcpapertype{system}

\usepackage{amsmath}
\usepackage{flushend}
\usepackage{placeins}
\usepackage{amssymb}
\usepackage{arydshln}
\usepackage{graphicx}
\usepackage{xspace}
\usepackage{makecell}
\usepackage{graphicx}
\usepackage{multirow}
\usepackage{listings}
\usepackage[caption=false]{subfig}
\usepackage[altpo,tsrm]{backnaur}
\usepackage[bookmarks,backref=true,linkcolor=black]{hyperref} 
\hypersetup{
  pdfauthor = {},
  pdftitle = {},
  pdfsubject = {},
  pdfkeywords = {},
  colorlinks=true,
  linkcolor= black,
  citecolor= black,
  pageanchor=true,
  urlcolor = black,
  plainpages = false,
  linktocpage
}

\DeclareRobustCommand{\sys}{\textsf{\mbox{Kyrix-S}}\xspace}
\DeclareRobustCommand{\visname}{\mbox{SSV}\xspace}
\DeclareRobustCommand{\visnames}{\mbox{SSVs}\xspace}
\DeclareRobustCommand{\subhead}[1]{\noindent\textbf{#1}}

\usepackage{xcolor}
\definecolor{mred}{rgb}{.80,.12,.30}
\definecolor{grey}{rgb}{.5,.5,.5}

\newif\ifnotes
\notestrue

\usepackage[normalem]{ulem}
\usepackage{soul}
\usepackage{hyperref}
\usepackage{bm}
\usepackage[english]{babel}
\usepackage[utf8]{inputenc}
\usepackage[noend]{algpseudocode}
\newcommand{\remco}[1]{\ifnotes{\leavevmode\color{mred}{\bf (Remco: #1)}}\else{#1}\fi}

\let\origcite\cite
\renewcommand{\cite}[1]{\ifnotes\mbox{\origcite{#1}}\else \origcite{#1}\fi}

\title{\sys: Authoring Scalable Scatterplot Visualizations\\of Big Data}

\author{Wenbo Tao, Xinli Hou, Adam Sah, Leilani Battle, Remco Chang and Michael Stonebraker}

\abstract{Static scatterplots often suffer from the overdraw problem on big datasets where object overlap causes undesirable visual clutter. The use of zooming in scatterplots can help alleviate this problem. With multiple zoom levels, more screen real estate is available, allowing objects to be placed in a less crowded way. We call this type of visualization \textit{scalable scatterplot visualizations}, or \visname for short. Despite the potential of \visnames, existing systems and toolkits fall short in supporting the authoring of \visnames due to three limitations. First, many systems have limited scalability, assuming that data fits in the memory of one computer.  Second, too much developer work, e.g., using custom code to generate mark layouts or render objects, is required. Third, many systems focus on only a small subset of the \visname design space (e.g. supporting a specific type of visual marks). To address these limitations, we have developed \sys, a system for easy authoring of \visnames at scale. \sys derives a declarative grammar that enables specification of a variety of \visnames in a few tens of lines of code, based on an existing survey of scatterplot tasks and designs. 
The declarative grammar is supported by a distributed layout algorithm which automatically places visual marks onto zoom levels. 
We store data in a multi-node database and use multi-node spatial indexes to achieve interactive browsing of large \visnames. 
Extensive experiments show that 1) \sys enables interactive browsing of \visnames of billions of objects, with response times under 500ms and 2) \sys achieves 4X-9X reduction in specification compared to a state-of-the-art authoring system.
} 


\CCScatlist{ 
 \CCScat{K.6.1}{Management of Computing and Information Systems}%
{Project and People Management}{Life Cycle};
 \CCScat{K.7.m}{The Computing Profession}{Miscellaneous}{Ethics}
}

\teaser{
  \centering
  \includegraphics[width=0.95\linewidth]{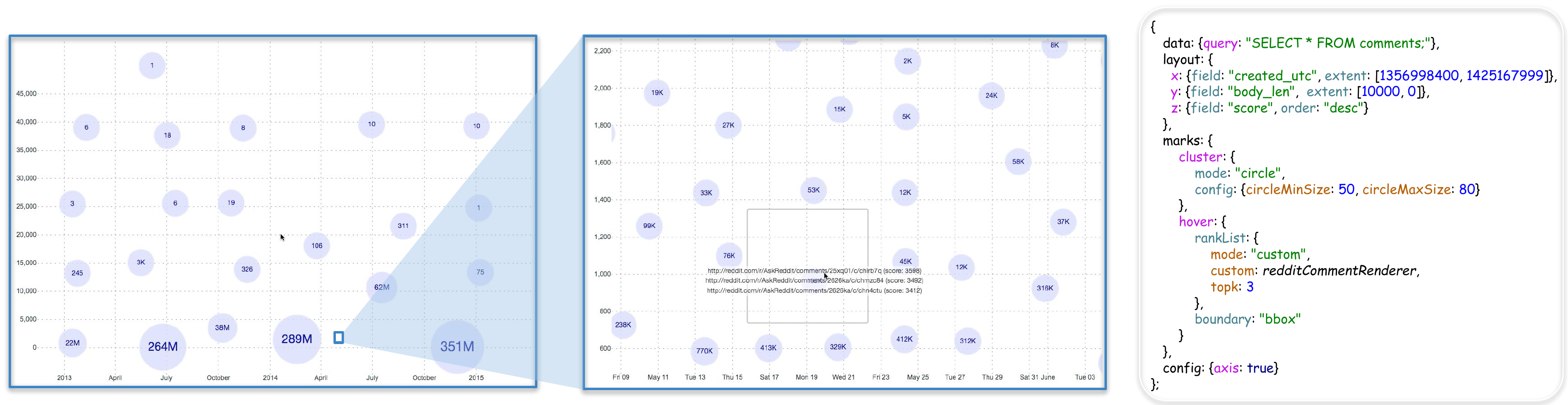}
  \vspace{-.75em}
  \caption{A scalable scatterplot visualization created by \sys and its \sys specifications. One billion comments made by users on \texttt{Reddit.com} from Jan 2013 to Feb 2015 are visualized on 15 zoom levels. On every level, $X$ and $Y$ axes are respectively the posting time and length of the comments. Each circle represents a cluster of comments. The number inside each circle is the size of the cluster and also encodes the radius of the circle. Using pan or zoom, the user can get either an overview (left) or inspect an area of interest (middle). One can hover over a circle to see three highest-scored comments in the cluster, as well as a bounding box showing the boundary of the cluster. \label{fig:teaser}}\label{fig:teaser}
}

\vgtcinsertpkg

\begin{document}



\maketitle

\section{Introduction}\label{sec:introduction}

Scatterplots are an important type of visualization used extensively in data science and visual analytic systems. Objects in a dataset are visualized on a 2D Cartesian plane, with the dimensions being two quantitative attributes from the objects. Each object can be represented as a point, polygon or other mark. Aggregation-based marks (e.g. pie chart, heatmap) can also be used to represent groups of objects. The user of a scatterplot can perform a variety of tasks to provide insights into the underlying data, such as discovering global trends, inspecting individual objects or characterizing distributions\cite{sarikaya2017scatterplots}. 

Despite the usefulness of static scatterplots, they suffer from significant overdraw problem on big datasets\cite{perrot2015large,mayorga2013splatterplots}. Here, we focus on scatterplots with millions to billions of objects, where significant overlap of marks is unavoidable, making the visualization ineffective. To address this issue in scatterplots, there has been substantial research\cite{heimerl2018visual,jo2018declarative,liu2013immens,lins2013nanocubes} on devising aggregation-based scatterplots using visual aggregates such as contours or hexagon bins. While avoiding visual clutter, these approaches do not support inspecting individual objects, which is a fundamental scatterplot task\cite{sarikaya2017scatterplots}. Prior works also used transparency\cite{fekete2002interactive,kosara2002focus+}, animation\cite{chen2018using} and displacements of objects\cite{waldeck2004mobile,keim1998gridfit,trutschl2003intelligently} to ease the overdraw problem. However, due to limited screen resolution, these methods have scalability limits.

On the other hand, the use of zooming in scatterplots has the potential to effectively mitigate visual clutter. By expanding the 2D Cartesian plane into a series of zoom levels with different scales, more screen resolution becomes available, allowing for object layouts that avoid occlusion and excessive density. Inspecting large amounts of objects thus becomes feasible. Aggregation-based marks such as circles or heatmaps can still be used to visualize groups of objects. 
Figure \ref{fig:teaser} shows such a visualization created by the system we introduce in this paper, which shows one billion comments made by users on \texttt{Reddit.com}, where $X$ is the posting time and $Y$ is the number of characters in the comments. Additional examples are in Figure \ref{fig:gallery}. For simplicity, we term such visualizations \textit{scalable scatterplot visualizations}, or \visname. 

There has been significant work on building systems/toolkits to aid the creation of \visnames (e.g.\cite{tao2019kyrix,bederson1994pad++,guo2018efficient,das2012efficient}). 
Specifically, prior systems can be classified into two categories: \textit{general pan/zoom systems} and \textit{specialized \visname systems}.
General pan/zoom systems are typically expressive, supporting not only \visnames, but also pan/zoom visualizations of other types of data (e.g. hierarchical and temporal data) or that connect multiple 2D semantic spaces\footnote{A 2D semantic space consists of zoom levels sharing the same coordinate system and visualizing the same type of objects. An \visname has only one semantic space. General pan/zoom systems typically allow ``semantic jumping'' from one semantic space to another\cite{tao2019kyrix,pietriga2005toolkit} (e.g. from a space of Reddit comments to a space of Reddit forums). }.
Specialized \visname systems (e.g. \cite{guo2018efficient,chen2014visual}), on the other hand, generally have a narrow focus on \visnames.  

While these systems have been shown to be effective, they can suffer from some drawbacks that limit their ability to support general \visname authoring at scale. In particular, \textbf{limited scalability} is a common drawback of both types of systems. As often as not, implementations assume all objects reside in the main memory of a computer\cite{chen2014visual, mayorga2013splatterplots, liao2017cluster, guo2018efficient,lins2013nanocubes,miranda2017topkube,bederson1994pad++,pietriga2005toolkit}. 

General pan/zoom systems, while being flexible, generally incur \textbf{too much developer work} due to their low-level nature. When authoring an \visname, the developer needs to manually generate the layout of visual marks on zoom levels. In very large datasets, there will be many levels (e.g. Google Maps has 20). Individually specifying the layout of a set of levels is tedious and error-prone. In particular, big or skewed data can make it challenging for the developer to specify a layout that avoids occlusion and excessive density in the visualization.

Another drawback of specialized \visname systems is \textbf{low flexibility}.
Oftentimes systems are hardcoded for specific scenarios (e.g., supporting specific types of visual marks such as heatmaps\cite{perrot2015large,lins2013nanocubes} or points\cite{das2012efficient,chen2014visual}, enforcing a density budget but not removing overlap, etc.) and are not extensible to general use cases. The developer cannot make free design choices when using these systems, and is forced to constantly switch tools for different application requirements.


\vspace{.2em}
In this paper, we describe \sys\footnote{The birth of \sys is driven by the limitations we see when we use Kyrix\cite{tao2019kyrix}, a general pan/zoom system we have developed, to build real-world \visname-based applications. The name \sys here suggests that we implement \sys as an extension of Kyrix for \visnames, rather than a replacement. \textsf{S} may suggest scale, scatterplots or spatial partitioning. More detailed discussion on the relationship between the two systems can be found in Sections \ref{sec:relatedwork} and \ref{sec:impl}. }, a system for \visname authoring at scale which addresses all issues of existing systems. 
To enable rapid authoring, we present a high-level declarative grammar for \visnames. We abstract away low-level details such as rendering of visual marks so that the developer can author a complex \visname in a few tens of lines of JSON. We show that compared to a state-of-the-art system, this is 4X--9X reduction in specification on several examples. In addition, we build a gallery of \visnames to show that our grammar is expressive and that the developer can easily extend it to add his/her own visual marks. 

This grammar for \visnames is supported by an algorithm that automatically chooses the layout of visual marks on all zoom levels, thereby freeing the developer from writing custom code.  
We store objects in a multi-node parallel database using multi-node spatial indexing. As we show in Section \ref{sec:evaluation}, this allows us to respond to any pan/zoom action in under 500ms on datasets with billions of objects.


To summarize, we make the following contributions:
\begin{itemize}
\vspace{-.75em}
    \item An integrated system called \sys for declarative authoring and rendering of \visnames at scale.\footnote{Code available at \url{https://github.com/tracyhenry/kyrix}}
    \vspace{-.9em}
    \item A concise and expressive declarative grammar for describing \visnames (Section \ref{sec:grammar}).
    \vspace{-.9em}
    \item A framework for offline database indexing and online serving that enables interactive browsing of large \visnames (Sections \ref{sec:sysarch} and \ref{sec:algorithm}). 
\vspace{-.5em}
\end{itemize}

\section{Related Works} \label{sec:relatedwork}

\subsection{General Pan/zoom Systems}\label{subsec:related_general}
A number of systems have been developed to aid the creation of general pan/zoom visualizations\cite{bederson1994pad++,bederson2003jazz,pietriga2005toolkit,tao2019kyrix}. These systems are expressive and capable of producing not only \visnames, but also pan/zoom visualizations of other types of data (e.g. hierarchical, temporal, etc) or with multiple semantic spaces connected by semantic zooms\cite{pietriga2005toolkit}. However, as mentioned in the introduction, these systems fall short in supporting \visnames due to \textbf{limited scalability} and \textbf{too much developer work}. 


Kyrix\cite{tao2019kyrix} is a general pan/zoom system we have developed. Here, we summarize the novel aspects of \sys compared to Kyrix and similar systems (e.g., ForeCache~\cite{battle2016dynamic}, Nanocubes~\cite{lins2013nanocubes}, imMens~\cite{liu2013immens}):
\begin{itemize}
\vspace{-.5em}
    \item \sys provides a high-level grammar for \visnames, which enables much shorter specification than what Kyrix's grammar requires for the same \visname (see Section \ref{subsec:authoringexp} for an empirical comparison);
\vspace{-.5em}
    \item \sys implements a layout generator which frees the developer from deciding the layout of objects on zoom levels. Kyrix does not assist the developer in choosing an object layout, which makes authoring \visnames using Kyrix fairly challenging;
\vspace{-.5em}
    \item  \sys is integrated with a distributed database which scales to billions of objects. In contrast, Kyrix only works with a single-node database which cannot scale to billions of objects. 
\vspace{-.5em}
\end{itemize}

Note that \sys has a narrow focus on \visnames and is not intended to completely replace general pan/zoom systems. As we will discuss more in Section \ref{sec:impl}, we implement \sys as an extension to Kyrix. 

\subsection{Specialized \visname Systems}\label{subsec:related_specialized}
There has been considerable effort made to develop specialized \visname systems, which mainly suffer from two limitations: \textbf{low flexibility} and \textbf{limited scalability}. 

Many systems focus on a small subset of the \visname design space, and are not designed/coded to be easily extensible. For example, many focus on specific visual marks such as small-sized dots (e.g.\cite{das2012efficient, chen2014visual, kefaloukos2014declarative}), heatmaps (e.g.\cite{perrot2015large,lins2013nanocubes, pahins2016hashedcubes,miranda2017topkube,liu2019smartcube}), text\cite{cartolabe}, aggregation-based glyphs\cite{liao2017cluster, beilschmidt2017linear} and contours\cite{mayorga2013splatterplots}. Some works maintain a visual density budget\cite{das2012efficient,perrot2015large,guo2018efficient}, while some focus on overlap removal\cite{beilschmidt2017linear,chen2014visual,derthick2003constant}. 
In contrast to these systems, \sys aims at a much larger design space. We provide a diverse library of visualization templates that are suitable for a variety of scatterplot tasks. For high extensibility, \sys's declarative grammar is designed with extensible components for authoring custom visual marks. 


In addition to the limited focus, most specialized \visname systems cannot scale to large datasets with billions of objects due to an in-memory assumption\cite{leafletcluster,delort2010vizualizing,guo2018efficient,mayorga2013splatterplots,chen2014visual,lekschas2019pattern,drosou2012disc,nutanong2012multiresolution}. We are only aware of the work by Perrot et al.\cite{perrot2015large} which renders large heatmaps using a distributed computing framework. However, that work only focuses on heatmaps. 

\vspace{.3em}
Specialized \visname systems generally come with a layout generation module which computes the layout of visual marks on each zoom level. The design of \sys's layout generation is inspired by many of them and bears similarities in some aspects. For example, favoring placements of important objects on top zoom levels is adopted by many works\cite{guo2018efficient,cartolabe,das2012efficient}. The idea of enforcing a minimum distance between visual marks comes from blue-noise sampling strategies\cite{perrot2015large,chen2014visual,guo2018efficient}.

However, the key differentiating factor of \sys comes from its more stringent requirements on scalability and the design space. These requirements (see Section \ref{sec:overview}) pose new algorithmic challenges. For instance, Sarma et al.\cite{das2012efficient} uses an integer programming solution without considering overlaps of objects. To enable overlap removal, one needs to add $O(n^2)$ pairwise non-overlap constraints into the integer program, making it hard to solve in reasonable time. 
As another example, Guo et al.\cite{guo2018efficient} and Chen et al.\cite{chen2014visual} do not support visual marks that show a group of objects with useful aggregated information. This requires a bottom-up aggregation process which breaks their top-down algorithmic flow. In order to scale to billions of objects, \sys cannot rely on existing algorithms and instead needs to compute visual mark layouts in parallel using a distributed algorithm as described in Section \ref{sec:algorithm}.


\subsection{Static Scatterplot Designs}
Alleviating the overdraw problem of static scatterplot visualizations has been a popular research topic for a long time. Many methods have been proposed, including binned aggregation\cite{moritz2019falcon,liu2013immens,jo2018declarative}, appearance optimization\cite{fekete2002interactive,kosara2002focus+,chen2018using}, data jittering\cite{waldeck2004mobile,keim1998gridfit,trutschl2003intelligently} and sampling\cite{dix2002chance,chen2019recursive}. We refer interested readers to existing surveys on scatterplot tasks and designs\cite{sarikaya2017scatterplots}, binned aggregation\cite{heimerl2018visual} and visual clutter reduction\cite{ellis2007taxonomy,elmqvist2009hierarchical}. \sys's design follows many guidelines in these works, which we elaborate in Section \ref{sec:overview}. 

\subsection{Declarative Visualization Grammars}\label{subsec:related_grammar}
Numerous declarative grammars have been proposed for authoring visualizations at different levels of abstractions. The first of these is Wilkinson's grammar of graphics (GoG)\cite{wickham2010layered}, which forms the basis of subsequent works. For example, ggplot2\cite{Wickham_2009} is the direct implementation of GoG in R and is widely used. D3\cite{d3} and Protovis\cite{bostock2009protovis} are low-level libraries that provide useful primitives for authoring basic visualizations. Vega is the first grammar that concerns specifications of interactions. Built on top of Vega, Vega-lite\cite{satyanarayan2017vegalite} offers a more succinct grammar for authoring interactive graphics. Recently, more specialized grammars have emerged for density maps\cite{jo2018declarative}, unit visualizations\cite{park2017atom}, and pan/zoom visualizations\cite{tao2019kyrix}. 

Despite the diversity of this literature, not many grammars support \visnames well. Some low-level grammars such as D3\cite{d3}, Vega\cite{vega} and Kyrix\cite{tao2019kyrix} can express \visnames, but the specification is often verbose due to their low-level and general-purpose nature. \sys, on the contrary, uses a high-level grammar that abstracts away unimportant low-level details. For example, switching mark representations can be simply done by changing a renderer type parameter (e.g. from ``circle'' to ``heatmap'') without writing a renderer. Furthermore, different from aforementioned grammars, \sys's grammar allows specifications of multiple zoom levels altogether with convenient components for specifying sampling/aggregation semantics. 


\section{Design Goals} \label{sec:overview}


Limitations of prior art, existing guidelines and our experience with \visname users drive the design of \sys. Here, we present a few goals we set out to achieve. 




\vspace{.3em}
\subhead{G1. Rapid authoring}. Our declarative grammar should enable specification of \visnames in a few tens of lines of code. This goal is inspired by the design rationale of several high-level declarative languages (e.g. Vega-lite\cite{satyanarayan2017vegalite} and Atom\cite{park2017atom}), and driven by the limitations we see in using Kyrix\cite{tao2019kyrix} to author \visnames. 

\vspace{.3em}
\subhead{G2. Visual expressivity}. \sys should enable the exploration of a broad \visname design space and not limit itself to specific visual representations. Moreover, it is crucial to allow inspection of individual objects in addition to showing aggregation information. As outlined by Sarikaya et al.~\cite{sarikaya2017scatterplots}, there are four common object-centric scatterplot tasks: \textit{identify object}, \textit{locate object}, \textit{verify object} and \textit{object comparison}. A recent study\cite{lekschas2019pattern} also highlights the importance of browsing objects in multi-scale visualizations.
%

\vspace{.3em}
\subhead{G3. Usable \visnames}. The \visnames authored with \sys should be usable, e.g. free of visual clutter, using simple visual aggregates, etc. We identify usability guidance from a range of surveys and \visname systems (e.g.\cite{elmqvist2009hierarchical,guo2018efficient,das2012efficient}), which we formally describe in Section \ref{sec:algorithm}. 



\vspace{.3em}
\subhead{G4. Scalability}. \sys should be able to handle large datasets with billions of objects and potentially skewed spatial distribution. This goal has the following two subgoals:
\begin{itemize}
\vspace{-.75em}
    \item \textbf{G4-a. Scalable offline indexing}. Offline indexing should finish in reasonable time on big data, and scale well as the data size grows. 
\vspace{-.75em}    
    \item \textbf{G4-b. Interactive online serving}. The end-to-end response time to any user interaction (pan or zoom) should be under 500ms, an empirical upper bound that ensures fluid interactions\cite{liu2014effects}. 
\end{itemize}
\vspace{-.75em}

In the rest of the paper, we justify the design choices we make by referencing the above goals when appropriate. 

\section{Declarative Grammar} \label{sec:grammar}
In this section, we present \sys's declarative grammar. We start with showing a gallery of example \visnames authored with \sys (Section \ref{subsec:gallery}), which we then use to illustrate the design of the grammar in Section \ref{subsec:grammarspec}.
\begin{figure*}[t]
\centering
\includegraphics[width=1\textwidth]{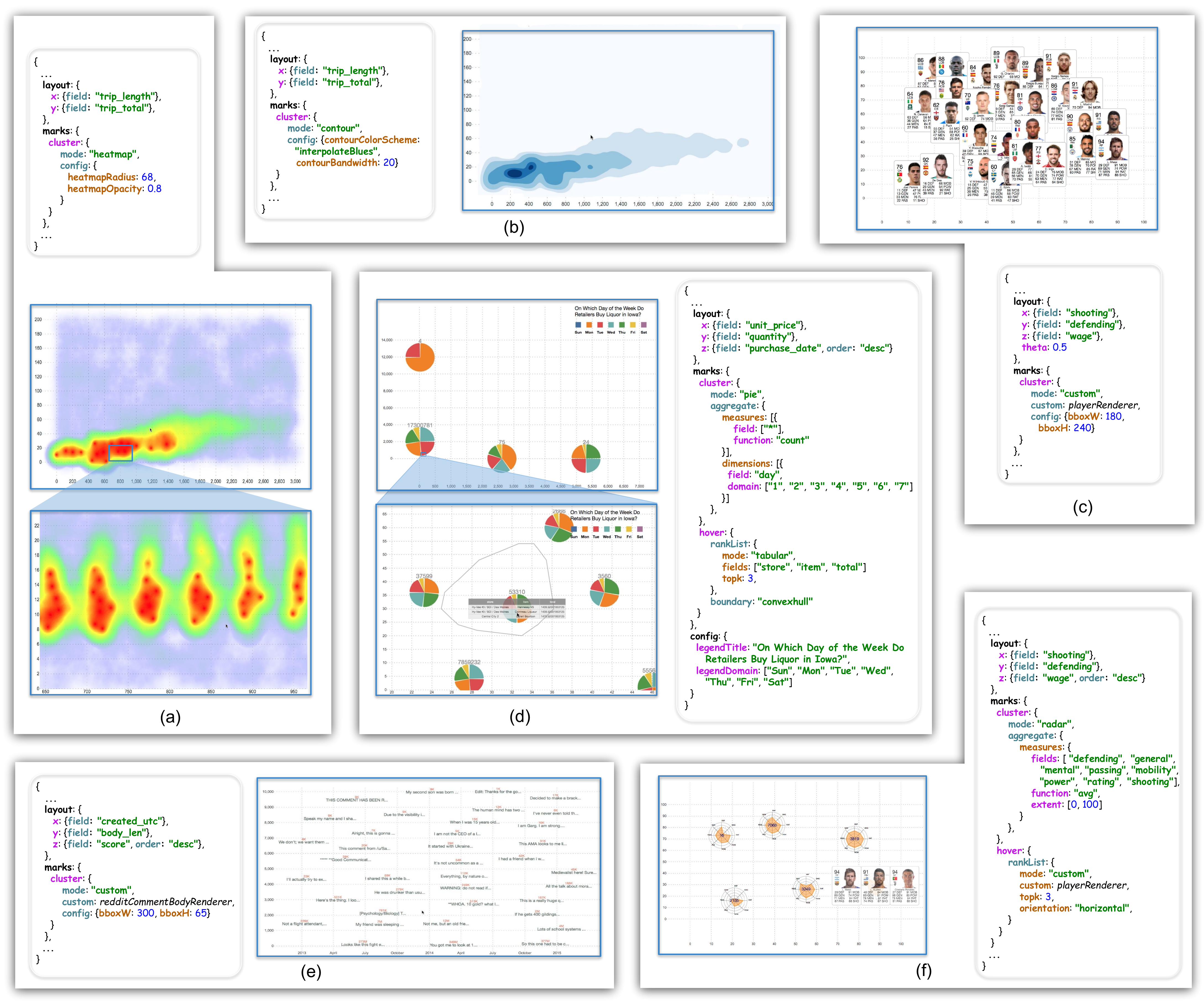}
\vspace{-2em}
\caption{A gallery of \visnames authored with \sys and their specifications. (a): a heatmap of 178.3 million taxi trips in Chicago since 2013, $X$: trip length (seconds), $Y$: trip total (dollars); (b): the same dataset/axes as (a) in contour plots; (c): an \visname of 18,207 soccer players in the video game FIFA19, $X$: shooting rating, $Y$: defense rating, $Z$: wage (i.e. highly-paid players appear on top zoom levels); (d): a pie-based \visname of 17.3 million liquor purchases by retailers in Iowa, $X$: unit price (dollars), $Y$: quantity (\# of bottles), $Z$: purchase date; (e): a text visualization of the dataset of one billion Reddit comments in Figure \ref{fig:teaser} with the same axes; (f): the same dataset/axes as (c) in a radar-chart.   \label{fig:gallery}}
\vspace{-1.5em}
\end{figure*}

\subsection{Example \visnames} \label{subsec:gallery}
Figure \ref{fig:gallery} shows a gallery of \visnames and their specifications. 

\vspace{.3em}
\subhead{Taxi}. In Figure \ref{fig:gallery}a, a multi-scale heatmap shows 178.5M taxi trips in Chicago since 2013, where $X$ is trip length (in seconds) and $Y$ is trip total (in dollars). In the overview (upper), the long thin ``heat'' region suggests that most trips have a similar total-length ratio. In a zoomed-in view (lower), we see vertical ``heat'' regions around entire minutes. In fact, more than 70\% of the trips have a length of entire minutes, indicating the possible prevalent use of minute-precision timers. Figure \ref{fig:gallery}b is the same representation of this dataset in contour lines. 

\vspace{.3em}
\subhead{FIFA}. The \visname in Figure \ref{fig:gallery}c visualizes 18,207 soccer players in the video game FIFA 19. $X$ and $Y$ are respectively the shooting and defensive rating of players. Players with the highest wages are shown at top levels. Lesser-paid players are revealed as one zooms in.
Figure \ref{fig:gallery}f is a radar-based \visname with the same $X$ and $Y$. 
Each radar chart shows the averages of eight ratings (e.g. passing, power) of a cluster of players. When hovering over a radar, three players from that cluster with the highest wages are shown. 

\vspace{.3em}
\subhead{Liquor}. Figure \ref{fig:gallery}d is an \visname of 17.3M liquor purchases by retailers in Iowa since 2012. $X$ and $Y$ axes are the unit price (dollars) and quantity (\# of bottles) of the purchases. Each pie shows a cluster of purchases grouped by day of the week. One can hover over a pie to see a tabular visualization of the three most recent purchases, as well as a convex hull showing the boundary of the cluster. \label{fig:runningexamplescreenshot}

\vspace{.3em}
\subhead{Reddit}. Figure \ref{fig:gallery}e is another representation of the one-billion Reddit comments dataset. Different from Figure \ref{fig:teaser}, comments are directly visualized as non-overlapping texts. The number above each comment represents how many comments are nearby, giving the user an understanding of the data distribution hidden underneath. 



\subsection{Grammar Design} \label{subsec:grammarspec}
\begin{figure}[t]
    \centering
    {\footnotesize
    \begin{bnf}
      \bnfprod{\visname}
        { \bnfpn{Marks}\bnfpn{Layout} \bnfpn{Data} \bnfpno{Config}}\\ 
      &&\hspace*{-100pt}\text{\textbf{; marks}} \nonumber\\
      \bnfprod{Marks}{\bnfpn{Cluster}\bnfpno{Hover}}\\
      \bnfprod{Cluster}{\bnfpn{Mode}\bnfpn{Aggregate}\bnfpno{Config}}\\
      \bnfprod{Hover}{\bnfpn{Ranklist}\bnfpn{Boundary}\bnfpno{Config}}\\
      \bnfprod{Mode}{\bnftd{Circle}\bnfor{}\bnftd{Contour}\bnfor{}\bnftd{heatmap}\bnfor{}}\nonumber\\
      \bnfmore{\bnftd{Radar}\bnfor{}\bnftd{Pie}\bnfor{}\bnfpn{Custom}}\\
      \bnfprod{Aggregate}{\bnfpn{Dimension}*\bnfpn{Measure}+}\\
      \bnfprod{Ranklist}{\bnfpn{Topk}(\bnftd{Tabular}\bnfor{}\bnfpn{Custom})}\\
      \bnfprod{Boundary}{\bnftd{Convex Hull}\bnfor{}\bnftd{BBox}}\\
      \bnfprod{Custom}{\bnfts{Custom JS mark renderer}}\\
      \bnfprod{Dimension}{\bnfpn{Field}\bnfpno{Domain}}\\
      \bnfprod{Measure}{\bnfpn{Field}\bnfpn{Function}\bnfpno{Extent}}\\
      \bnfprod{Topk}{\bnfts{A positive integer}}\\
      \bnfprod{Domain}{\bnfts{A list of string values}}\\
      \bnfprod{Function}{\bnftd{Count}\bnfor{}\bnftd{Sum}\bnfor{}\bnftd{Avg}\bnfor{}\bnftd{Min}\bnfor{}}\nonumber\\
      \bnfmore{\bnftd{Max}\bnfor{}\bnftd{Sqrsum}}\\
      &&\hspace*{-100pt}\text{\textbf{; layout}} \nonumber\\
      \bnfprod{Layout}{\bnfpn{X}\bnfpn{Y}\bnfpn{Z}\bnfpno{Theta}}\\
      \bnfprod{X}{\bnfpn{Field}\bnfpno{Extent}}\\
      \bnfprod{Y}{\bnfpn{Field}\bnfpno{Extent}}\\
      \bnfprod{Z}{\bnfpn{Field}\bnfpn{Order}}\\
      \bnfprod{Theta}{\bnfts{A number between 0 and 1}}\\
      \bnfprod{Field}{\bnfts{A database column name}}\\
      \bnfprod{Extent}{\bnfts{A pair of float numbers}}\\
      \bnfprod{Order}{\bnftd{Ascending}\bnfor{}\bnftd{Descending}}\\
      &&\hspace*{-100pt}\text{\textbf{; data}} \nonumber\\
      \bnfprod{Data}{\bnfts{a database query}}\\
      &&\hspace*{-100pt}\text{\textbf{; config}} \nonumber\\
      \bnfprod{Config}{\bnfts{Key value pairs}}
    \end{bnf}
    }
    \vspace{-1.5em}
    \caption{\sys's declarative grammar in the BNF notation. Inside $\langle \rangle$ or [] is a component. Every rule (1-24) defines what the left-hand side component is composed of. On the right hand side of a rule, $|$ means OR, * means zero or more, + means one or more and [] means that a component is optional. } 
    \vspace{-1.5em}
    \label{fig:grammar}
\end{figure}

The primary goal of \sys's declarative grammar is to help the developer quickly navigate a large \visname design space (\textbf{G1} and \textbf{G2}). The high-level design of the grammar closely follows a survey of scatterplots designs and tasks by Sarikaya et al.\cite{sarikaya2017scatterplots}, which outlined four common design variables of scatterplot visualizations: 
\textit{point encoding} (i.e. visual representation of one object), \textit{point grouping} (i.e. visual representation of a group of objects), \textit{point position} (e.g. subsampling, zooming) and \textit{graph amenities} (e.g. axes, annotations). These design variables map to the highest-level components in \sys's grammar, i.e., \textit{Marks}, \textit{Layout}, \textit{Data} and \textit{Config}, as illustrated in Figure \ref{fig:grammar} using the BNF notation\cite{knuth1964backus}. We elaborate the design of them in the following.



\subsubsection{Marks: Templates + Extensible Components}
The \textit{Marks} component (Rules 2-14\footnote{Hereafter, rules referenced inside parentheses implicitly refer to rules in Figure \ref{fig:grammar}. A rule defines the composition logic of one component in the grammar.}) defines the visual representation of one or more objects, and covers both \textit{point encoding} and \textit{point grouping} in \cite{sarikaya2017scatterplots}. 
Visual marks of a single or a cluster of objects span a huge space of possible visualizations. To keep our grammar high-level (\textbf{G1}), we adopt a \textit{templates}+\textit{extensible components} methodology, where we provide a diverse library of template mark designs, and offer extensible components for authoring custom marks. 





We divide the \textit{Marks} component into two subcomponents: \textit{Cluster} (Rule 3) and \textit{Hover} (Rule 4). 


\vspace{.3em}
\noindent \textbf{Cluster:} cluster marks are static marks rendering one or a group of objects. 
Currently, \sys has five built-in \textit{Cluster} marks including \textsc{Circle} (Figure \ref{fig:teaser}), \textsc{Contour} (Figure \ref{fig:gallery}b), \textsc{Heatmap} (Figure \ref{fig:gallery}a), \textsc{Radar} (Figure \ref{fig:gallery}f) and \textsc{Pie} (Figure \ref{fig:gallery}d). The developer can choose one of these marks by specifying just a name (\textbf{G1}).
These built-in \textit{Cluster} marks are carefully chosen to cover a range of aggregate-level \visname tasks\cite{sarikaya2017scatterplots}. For example, heatmaps and contour plots enable the user to \textit{characterize distribution} and \textit{identify correlation} between the two axes. The user can perform \textit{numerosity comparison} and \textit{identify anomalies} with circle-based \visnames. Radar-based and pie-based \visnames allow for \textit{exploring object properties within a neighborhood}. For fast authoring, \sys sets reasonable default values for many parameters (\textbf{G1}), e.g., inner/outer radius of a pie and bandwidth of heatmaps. The developer can also customize (\textbf{G2-b}) using a \textit{Config} component (Rules 3 and 24).


With the \textit{Custom} component (Rules 5 and 9), the developer can specify custom visual marks easily. For example, player profiles in Figure \ref{fig:gallery}c are specified as a custom visual mark. 
\sys currently supports arbitrary Javascript-based renderers (e.g. D3\cite{d3} or Vega-lite-js\cite{vegalitejs}). For increased expressivity, a custom mark renderer is passed all useful information about a cluster of objects, including aggregation information in both \textit{Aggregate} and \textit{Hover}. As an example, the custom renderer in Figure \ref{fig:gallery}e displays both an example comment and the size of the cluster. 
More importantly, \textit{Custom} also facilitates rapid future extension of \sys, allowing easy addition of built-in mark types. 

The \textit{Aggregate} component (Rule 6) specifies details of aggregations statistics shown by a \textit{Cluster} mark, and is composed of \textit{Dimension}s (Rule 10) and \textit{Measure}s (Rule 11). A \textit{Dimension} is a categorical field of the objects indicating how objects are grouped (e.g. by day of the week in Figure \ref{fig:gallery}d). A \textit{Measure} defines an aggregation statistic (e.g. average of a rating in Figure \ref{fig:gallery}f). Currently \sys supports six aggregation functions: count, average, min, max, sum and square sum (Rule 14). 

\vspace{.3em}
\noindent \textbf{Hover:} Hover marks add more expressivity into the grammar by showing additional marks when the user hovers over a \textit{Cluster} mark. For example, in Figure \ref{fig:teaser} three example comments are shown upon hovering a circle. The motivation for adding this component is two-fold. 

First, as outlined in \textbf{G2}, we want to enable tasks that require inspection of individual objects in addition to showing visual aggregates with \textit{Cluster} marks. To this end, we design a \textit{Ranklist} component which visualizes objects with top-k importance (Rule 7). The importance of objects is defined in the \textit{layout} component as a field from the objects. We offer a default tabular visualization template (e.g. Figure \ref{fig:gallery}d), and allow custom marks via \textit{Custom} (e.g. player profiles in Figure \ref{fig:gallery}f).

Secondly, multi-scale visualizations often suffer from the ``desert fog'' problem\cite{jul1998critical}, where the user is lost in the multi-scale space and not sure what is hidden underneath the current zoom level. \textit{Boundary} is designed to aid the user in navigating (\textbf{G3}) by showing the boundaries of a cluster of objects (Rule 8), using either the convex hull (Figure \ref{fig:gallery}d) or the bounding box (Figure \ref{fig:teaser}). By hinting that there is more to see by zooming in, more interpretability is added to the visualization.

\subsubsection{Layout: Configuring All Zoom Levels at Once}

The \textit{Layout} component (Rules 15-22) controls the placement of visual marks\footnote{For KDE-based \visnames (e.g. heatmaps and contours), a visual mark here refers to the kernel density estimates generated by a weighted object.} on zoom levels, which corresponds to the \textit{point position} design variable in \cite{sarikaya2017scatterplots}. We aim to assist the developer in specifying the layout for all zoom levels together rather than independently, motivated by the limitation of general pan/zoom systems\cite{tao2019kyrix,bederson2003jazz,bederson1994pad++} that mark placements are manually configured for every zoom level. 

$X$ and $Y$ (Rules 16 and 17) define the two spatial dimensions. The only specifications required are two raw data columns that map to the two dimensions (e.g. trip length and total in Figures \ref{fig:gallery}a and \ref{fig:gallery}b). An optional \textit{Extent} component (Rule 21) can be used to indicate the visible range of raw data values on the top zoom level. 


The $Z$ component (Rule 18) controls how visual marks are distributed across zoom levels. Drawn from prior works\cite{das2012efficient,guo2018efficient,cartolabe}, we use a usability heuristic that makes objects with higher importance more visible on top zoom levels. The importance is defined by a field of the objects. For example, in Figure \ref{fig:gallery}e, highest-scored comments are displayed on top zoom levels. 

Optionally, \textit{Theta} is a number between 0 and 1 indicating the amount of overlap allowed between \textit{Cluster} marks (Rule 19), with 0 being arbitrary overlap is allowed and 1 being overlap is not allowed. For instance, \textit{Theta} is 0.5 in Figure \ref{fig:gallery}c, making the player profiles overlap to a certain degree. 

The above layout-related parameters serve as inputs to the layout generator, which we detail in Section \ref{sec:algorithm}.


\subsubsection{Data and Config}

We assume that the raw spatial data exists in the database, and can be specified as a SQL query (Rule 23). 
The highest-level \textit{Config} component corresponds to the design variable \textit{graph amenities} in\cite{sarikaya2017scatterplots}. The developer can use it to specify global rendering parameters such as the size of the top zoom level, number of zoom levels, as well as annotations such as axes, grid lines and legends. 

\section{Optimization Framework} \label{sec:sysarch}
\begin{figure}[t]
\centering
\includegraphics[width=0.4\textwidth]{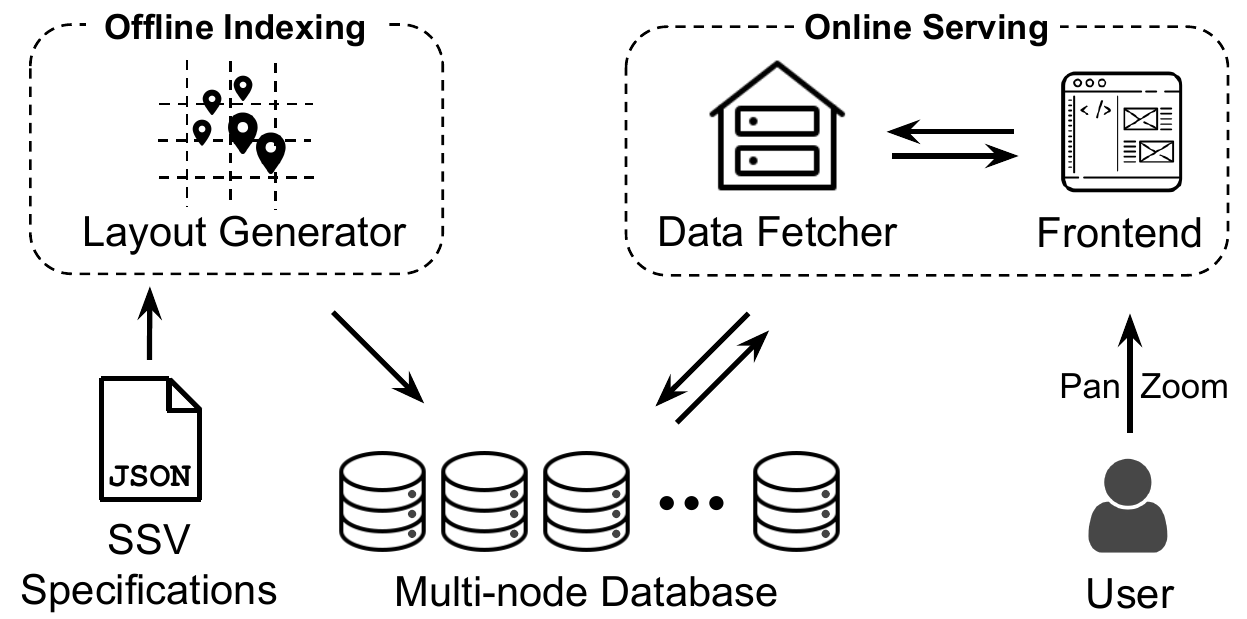}
\vspace{-.5em}
\caption{\sys~optimization framework. \label{fig:architecture}}
\vspace{-1.5em}
\end{figure}

Figure \ref{fig:architecture} illustrates the optimization framework adopted by \sys to scale to large datasets(\textbf{G4}). There are two main phases: offline indexing and online serving. 
Specifically, given an \visname specification, the layout generator computes offline the placement of visual marks on zoom levels using several usability considerations (\textbf{G3}), e.g., bounded visual density, free of clutter, etc. Along the way, useful aggregation information (e.g. statistics and cluster boundaries) is also collected. The computed layout information is stored in a multi-node database with multi-node spatial indexes. Online, the data fetcher communicates with the frontend and fetches data in user's viewport from the multi-node database with sub-500ms response times (\textbf{G4-b}). In the next section, we describe these two components in greater detail. 

\section{Layout Generation and Data Fetching} \label{sec:algorithm}



Here, we first describe how we model the layout generation problem (Section \ref{subsec:layoutproblemdef}). We then describe a single-node layout algorithm (Section \ref{subsec:singlenodealgo}), which is the basis of a distributed algorithm detailed in Section \ref{subsec:multinodealgo}. Lastly, Section \ref{subsec:datafetching} describes the design of the data fetcher.

\subsection{Layout Generation: Problem Definition} \label{subsec:layoutproblemdef}
We assume that there is a discrete set of zoom levels numbered 1, 2, 3$\ldots$ from top to bottom with a constant zoom factor between adjacent levels (e.g. 2 as in many web maps). The layout generation problem concerns how to, in a scalable manner, place visual marks onto these zoom levels in a general way that works for any \visname that \sys's declarative grammar can express (\textbf{G2}). 



To aid the formulation of the layout generation problem, we collect a set of existing layout-related usability considerations from prior \visname systems and surveys\cite{guo2018efficient,beilschmidt2017linear, chen2014visual,das2012efficient,elmqvist2009hierarchical,liao2017cluster}, and list them as subgoals of \textbf{G3: Usable \visnames}.

\vspace{.3em}
\subhead{G3-a. Non/partial overlap}.
\textit{Cluster} visual marks (Rule 3) should not overlap or only overlap to a certain degree (if specified by \textit{Theta} in Rule 19). For simplicity, we assume that \textit{Cluster} marks have a fixed-size bounding box, which is either decided by \sys or specified by the developer (see Figure \ref{fig:gallery}e for an example). We then only check the overlap of bounding boxes. 

\vspace{.3em}
\subhead{G3-b. Bounded visual density}. Mark density in any viewing region should not exceed an upper bound. Excessive density stresses the user and slows down both the client and the server. \sys sets a default upper bound $K$ on how many marks should exist in any viewport-sized region based on empirical estimates of the processing capability of the database and the frontend. 
We should also avoid very low visual density, which often leads to too many zoom levels and thus increased navigation complexity. We therefore try to maximize spatial fullness without violating the overlap constraint and the density upper bound. 


\vspace{.3em}
\subhead{G3-c. Zoom consistency}. If one object is visible on zoom level $i$, either through a custom \textit{Cluster} mark or a \textit{Ranklist} mark (Rule 7), it should stay visible on all levels $j > i$. This principle is adopted by many \visname systems that support inspection of individual objects (e.g. \cite{das2012efficient,chen2014visual,guo2018efficient}). The rationale is to aid object-centric tasks where keeping track of locations of objects is important. 

\vspace{.3em}
\subhead{G3-d. Data abstraction quality}. Data abstraction characterized by visual marks should be interpretable and not misinform the user. For \textit{Cluster} marks, it is important to reduce \textit{within-cluster variation}\cite{cui2006measuring,yang2003interactive,elmqvist2009hierarchical}, which can be characterized by average distance of objects to the visual mark that represent them\cite{cui2006measuring}. We also adopt an \textit{importance policy}, where objects with higher importance (Rule 18) should be more likely to be visible on top zoom levels. This is a commonly adopted principle to help the user see representative objects early on\cite{guo2018efficient,das2012efficient}. 


\vspace{.3em}
\subhead{Discussion}. Despite that subgoals \textbf{G3-a$\sim$d} are all from existing works, we are not aware of any prior system that addresses all of them. As mentioned in Section \ref{sec:relatedwork}, a key distinction of \sys's layout generation lies in the more stringent requirements of scalability and the design space. Due to this broad focus, finding an ``optimal layout'' with the objectives and constraints in \textbf{G3-a$\sim$d} is hard. In fact, a prior work\cite{das2012efficient} proves that with only a subset of \textbf{G3-a$\sim$d}, finding the optimal layout is NP-hard (for an objective function they define). Therefore, we do not attempt to define a formal constraint solving problem. Instead we keep our goals qualitative and look for heuristic solutions.


\subsection{A Single-node Layout Algorithm} \label{subsec:singlenodealgo}

Here, we describe a single-node layout algorithm which assumes that data fits in the memory of one computer. 

We assume that the $X$/$Y$ placement of a \textit{Cluster} mark comes from an object it represents. Alternatively, one could consider inexact placement of the marks (e.g. ``median location'' or binned aggregation), which we leave as our future work. Additionally, we assume that the $X$/$Y$ placement of a \textit{Hover} mark is the same as the corresponding \textit{Cluster} mark. So in the rest of Section \ref{sec:algorithm}, any mention of mark refers to a \textit{Cluster} mark if not explicitly stated. 

We make two important algorithmic choices. First, we enforce a minimum distance between marks in order to cope with the overlap and density constraints (\textbf{G3-a} and \textbf{G3-b}). Second, we use a hierarchical clustering algorithm to ensure zoom consistency (\textbf{G3-c}) and data abstraction quality (\textbf{G3-d}). 



\begin{figure}[t]
\centering
\includegraphics[width=0.4\textwidth]{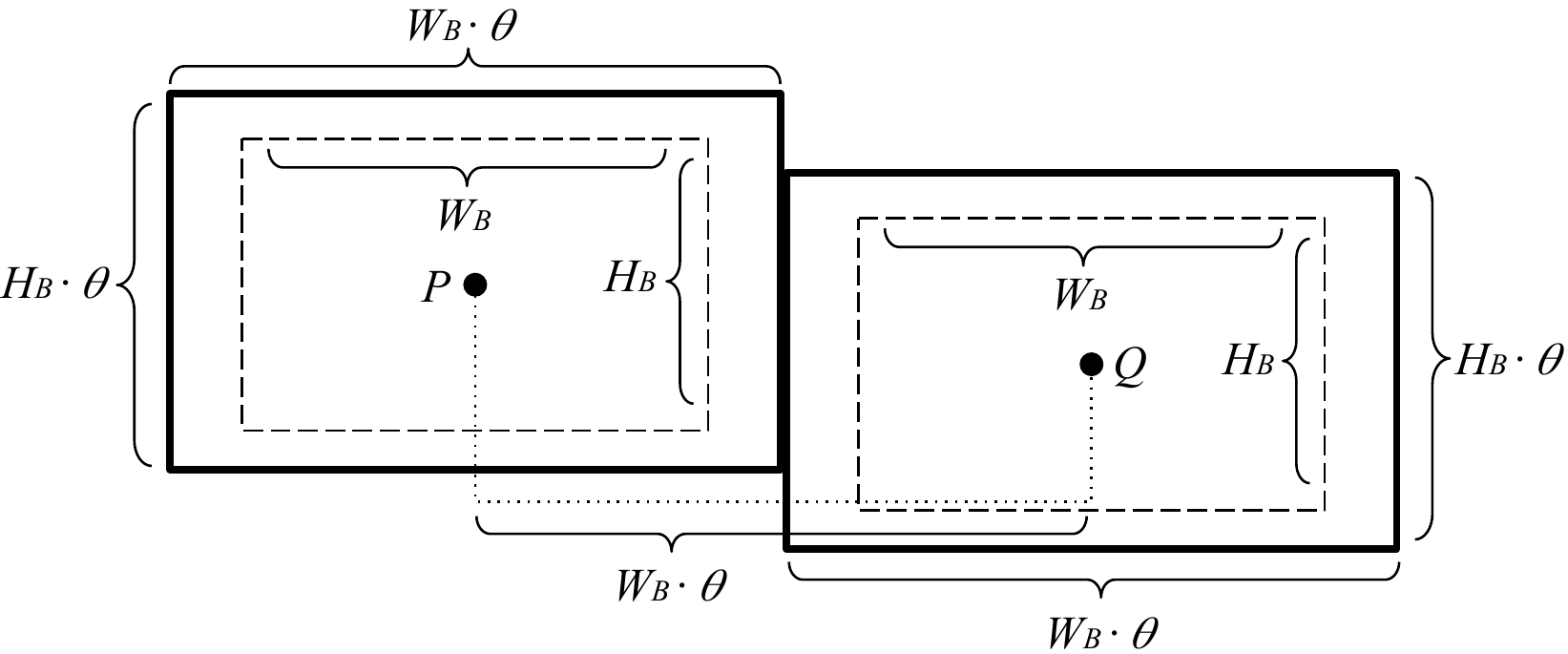}
\vspace{-1.5em}
\caption{Marks $P$ and $Q$ with an $ncd$ of $\theta$. Inner boxes (dashed) are the bounding boxes of the marks. Outer boxes (solid) are bounding boxes scaled by a factor of $\theta$. Scaled boxes do not overlap and touch on one side. In general, for any two marks that have $ncd$ greater than $\theta$, their bounding boxes do not overlap after being scaled by a factor of $\theta$. \label{fig:theta}}
\vspace{-1.25em}
\end{figure}

\vspace{.3em}
\subhead{Enforcing a minimum distance between marks}. For overlap and density constraints, we make use of the \textit{normalized chessboard distance} ($ncd$) between two marks $P$ and $Q$:
\begin{equation}
    ncd(P, Q) = \max(\frac{|P_x - Q_x|}{W_B}, \frac{|P_y-Q_y|}{H_B}) \nonumber
\end{equation}
where $P_x$($P_y$) is the $x$($y$) coordinate of the centroid of $P$ in the pixel space and $W_B$($H_B$) is the width (height) of the bounding box of a mark (note that bounding boxes of marks are of the same size). 

$ncd$ helps us reason about non/partial overlap constraints. If $ncd(P, Q)\geq1$, $P$ and $Q$ do not overlap because they are at least one bounding box width/height away on $X$ or $Y$. Even if $ncd$ is smaller than one, the degree of overlap is bounded. For example, if $ncd(P, Q)=0.5$, the centroids of $P$ and $Q$ remain visible despite the potential overlap. 

To this end, we set a lower bound $\theta$ on the $ncd$ between any two visual marks, which is specified through the \textit{Theta} component (e.g. Figure \ref{fig:gallery}c) or built-in with \textit{Cluster} marks.

We also use $\theta$ to enforce the visual density upper bound $K$ (\textbf{G3-b}). Intuitively, the smaller $\theta$ is, the closer marks are, and thus the denser the visualization is. We search for the smallest $\theta$ (for maximum spatial fullness, \textbf{G3-b}) that does not allow more than $K$ marks in any viewport-sized region ($W_V\times H_V$). To find this $\theta$ value, we show in Figure \ref{fig:theta} another perspective on how $\theta$ controls the placement of marks: enforcing that any $ncd\geq \theta$ is equivalent to scaling the bounding boxes of marks by a factor of $\theta$, and then enforcing that none of these scaled bounding boxes overlap. So we are left with a simple bin-packing problem. For a given $\theta$, the maximum number of marks that can be packed into a viewport is:
\begin{equation}
    \mathcal{P}(\theta)=\left\lceil\frac{W_V}{W_B\cdot\theta} \right\rceil \cdot \left\lceil  \frac{H_V}{H_B\cdot\theta} \right\rceil \nonumber
\end{equation}
With this, we can find the smallest $\theta$ such that $\mathcal{P}(\theta)\leq K$ using a binary search on $\theta$. 

We take the larger $\theta$ calculated/specified for the overlap and density constraints. By imposing this lower bound on $ncd$, these two constraints are strictly satisfied.


\begin{figure}[t]
\centering
\includegraphics[width=0.45\textwidth]{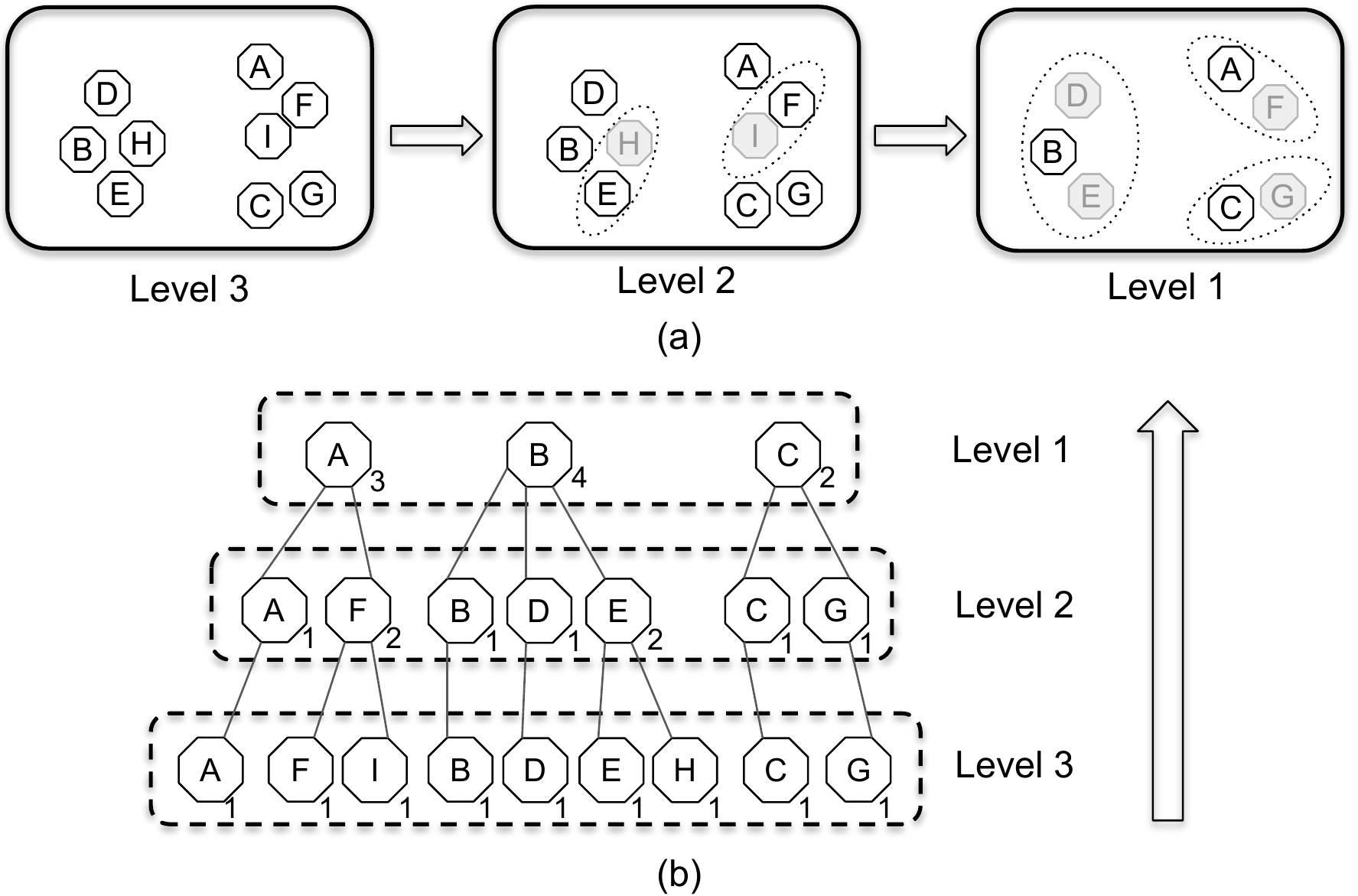}
\vspace{-.5em}
\caption{An illustration of the hierarchical clustering. There are 9 objects A-I, in decreasing order of importance. Each octagon is a cluster, with the representative object inside it. (a): Three zoom levels constructed. A dashed ellipse indicates the merging of the lighter cluster into the darker one. (b): A tree representation of the hierarchical clusters. The number next to a cluster is the number of objects this cluster represents. These numbers, along with other possible aggregation information, are computed when clusters merge. \label{fig:single_node}}
\vspace{-1.25em}
\end{figure}

\vspace{.3em}
\subhead{Hierarchical clustering}. The key part of the algorithm is a bottom-up hierarchical clustering process. Suppose there are $\eta$ zoom levels. We start with a fake bottom level $\eta + 1$ where every object is in its own cluster. Each cluster's aggregation information (e.g. aggregated stats and cluster boundaries) is initialized using the only object in it, which we call the ``representative object'' of a cluster in the following. 

Then we build the clusters level by level. For each zoom level $i\in[1, \eta]$, we construct a new set of clusters by merging the clusters on level $i+1$. Zoom consistency (\textbf{G3-c}) is then guaranteed because each zoom level merges clusters from the one level down. By mathematical induction, we can show that if an object is visible on level $i$, it is visible on any level $j>i$.

Specifically, we iterate over all clusters on level $i+1$ in the order of the importance of their representative objects, which is a greedy strategy to make important objects more visible (\textbf{G3-d}). For each cluster $\alpha$ on level $i+1$, we search for a cluster $\beta$ on the current level $i$ with the closest $ncd$. If this $ncd$ is smaller than $\theta$, we merge $\alpha$ into $\beta$; otherwise we add $\alpha$ to level $i$. By merging a cluster into its nearest neighbor (measured in $ncd$), within-cluster variances can be reduced (\textbf{G3-d}). Figure \ref{fig:single_node} shows an example with 9 objects and 3 zoom levels. 

\vspace{.3em}
\subhead{Identifying outliers}. The single-node algorithm preserves an outlier if it is not within $\theta$ $ncd$ of any other object. To identify less isolated outliers, one would need to assign to each object a score (i.e. the importance field) indicating how distant an object is from other objects. Kernel density estimations would be an example of such type of score.


\vspace{.3em}
\subhead{Optimizations and complexity analysis}. Let $n$ be the total number of objects. When constructing clusters for level $i$, sorting the clusters on level $i+1$ takes $O(n\log n)$. We maintain a spatial search tree (e.g. R-tree) of the clusters on level $i$ so that nearest neighbor searches can be done in $O(\log n)$. Inserting a new cluster into the tree also takes $O(\log n)$. Therefore, the overall time complexity of this algorithm is $O(n\log n)$ if we see the number of zoom levels as a constant. 



\subsection{A Multi-node Distributed Layout Algorithm} \label{subsec:multinodealgo}


The algorithm presented in Section \ref{subsec:singlenodealgo} only works on a single machine which has limited memory. Here, we extend it to work with a multi-node database system.\footnote{The distributed algorithm proposed here works with any multi-node database that supports basic data partitioning (e.g. Hash-based) and 2D spatial indexes.}

\begin{figure*}[t]
\centering
\includegraphics[width=0.88\textwidth]{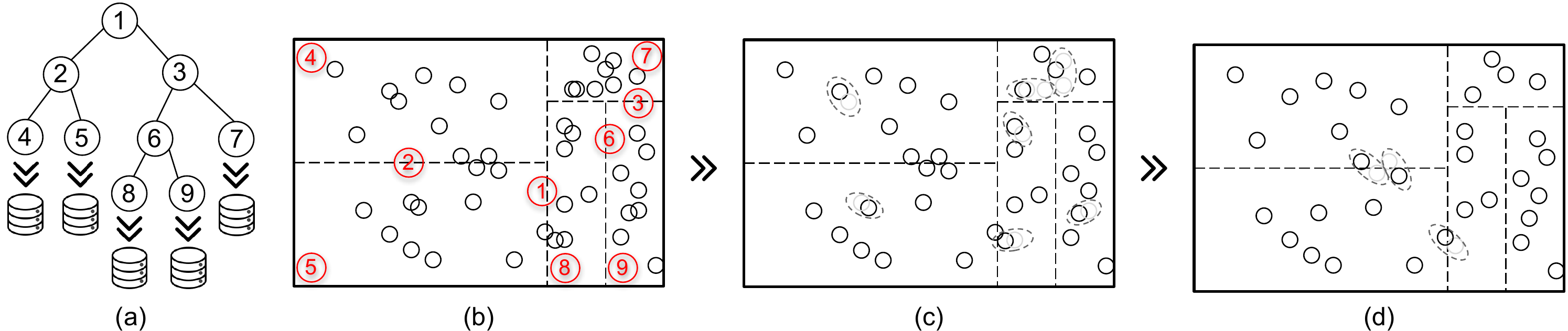}
\vspace{-1em}
\caption{An illustration of the distributed clustering algorithm for zoom level $i$. (a), (b): clusters on zoom level $i+1$ are spatially partitioned and stored on multiple database nodes. KD-tree is used for skew-resilient partitioning. In (a), non-leaf tree nodes (1, 2, 3 and 6) represent KD-tree splits, while leaf tree nodes (4, 5, 7, 8 and 9) correspond to actual partitions. Each circle in (b) is a mark/cluster; (c): the single-node algorithm is run for each partition in parallel, merging clusters that have an $ncd$ smaller than $\theta$; (d): merging clusters close to partition boundaries.
\label{fig:multinode}}
\vspace{-1em}
\end{figure*}

Given the sequential nature of the single-node algorithm, one major challenge here is how to utilize the parallelism offered by the multi-node database. 
Our idea is to spatially partition a zoom level, perform clustering in each partition independently in parallel and then merge the partitions. Figure \ref{fig:multinode} shows an illustration of the three steps. We detail them in the following, assuming the context of constructing clusters on zoom level $i$ from the clusters on level $i+1$.

\vspace{.3em}
\subhead{Step 1: skew-resilient spatial partitioning}. 
We use a KD-tree\cite{bentley1990k} to spatially partition the 2D plane so that each resulting partition has similar number of clusters from zoom level $i+1$. Note that each cluster belongs to exactly one partition according to its centroid. A KD-tree is a binary tree (Figure \ref{fig:multinode}a) where every non-leaf tree node represents a split of a subplane, and every leaf tree node is a final partition stored as a table in one database node. KD-tree splits are axis-aligned and alternate between horizontal and vertical as one goes down the hierarchy. For each split, the median value of the corresponding axis is used as the split point. We stop splitting when the number of clusters in a partition can fit into the memory of one database node. 

\begin{figure}[t]
\centering
\vspace{-.25em}
\includegraphics[width=0.4\textwidth]{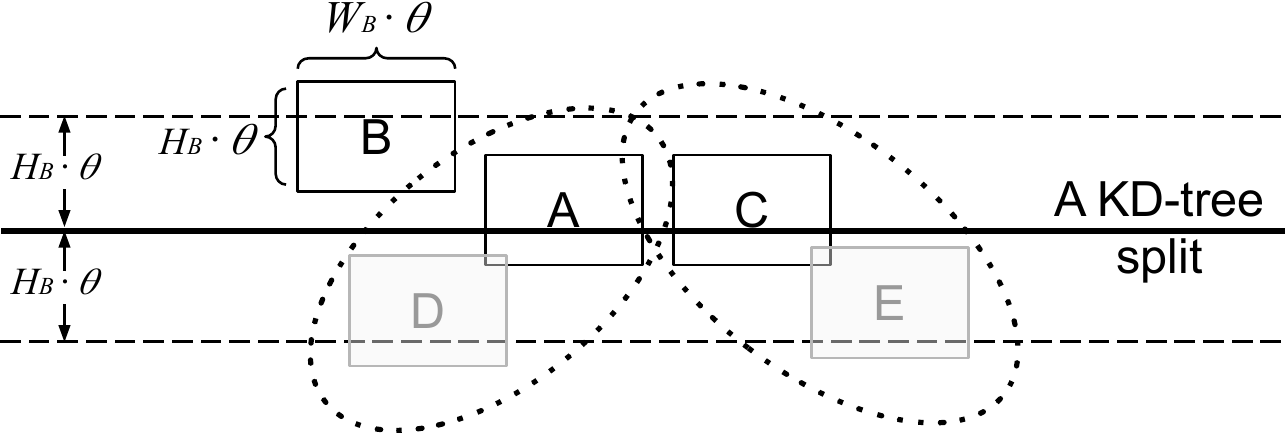}
\vspace{-1.25em}
\caption{An example of merging clusters along a KD-tree split. \label{fig:mergepartition}}
\vspace{-1.5em}
\end{figure}


\vspace{.3em}
\subhead{Step 2: processing partitions in parallel}. Since each partition fits in the memory of one database node, we can efficiently run the single-node clustering algorithm on each partition in parallel. As a result, a new set of clusters is produced in each partition where no two clusters have an $ncd$ smaller than $\theta$ (Figure \ref{fig:multinode}c). 

\vspace{.3em}
\subhead{Step 3: merging clusters on partition boundaries}. After Step 2, some clusters close to partition boundaries may have an $ncd$ smaller than $\theta$. Step 3 resolves these border cases by merging clusters along KD-tree splits. We ``process'' (i.e. merging clusters along) KD-tree splits in a bottom-up fashion, starting with splits that connect two leaf partitions. After the KD-tree root is processed, we finish the layout generation for level $i$. 



When processing a given split, we make use of the fact that only clusters whose centroid is within a certain distance to the split ($W_B\cdot \theta$ or $H_B\cdot \theta$ depending on the orientation of the split) need to be considered. Consider the horizontal split in Figure \ref{fig:mergepartition}. The two horizontal dashed lines indicate the range of cluster centroids that we need to consider. Any cluster whose centroid is outside this range is at least $\theta$ away (in $ncd$) from any cluster on the other side of the split. 

We use a greedy algorithm to process a KD-tree split. We iterate over all clusters in the aforementioned range in the order of their $x$ coordinates ($y$ if the split is vertical). 
We keep track of the last added/merged cluster $\alpha$. Let $\beta$ be the currently considered cluster. If $ncd(\alpha, \beta)\geq \theta$, we add $\beta$ and set $\alpha$ to $\beta$; otherwise we merge $\alpha$ and $\beta$. The one with the less important representative object is merged into the other (\textbf{g3-d}). Then we update $\alpha$ accordingly. 


Consider again Figure \ref{fig:mergepartition}. There are five clusters A-E in decreasing importance order. The boxes around clusters are their bounding boxes scaled by a factor of $\theta$. So if two boxes overlap, two corresponding clusters have an $ncd$ smaller than $\theta$ (see Figure \ref{fig:theta}). The above algorithm iterates over the clusters in the following order: $B, D, A, C, E$. When $\beta=A$, $\alpha=D$. $D$ is then merged into $A$ because $ncd(A, D) < \theta$ and $D$ has a less important representative object. For the same reason, $E$ is merged into $C$. 

\vspace{.3em}
\subhead{Optimizations and complexity analysis}. Let $M$ be the upper bound on the number of clusters that can fit in memory. Hence there are roughly $T=\frac{n}{M}$ partitions, which means there are $O(T)$ KD-tree nodes. Determining the splitting point can be done in $O(\log n)$, thus constructing the spatial partitions takes $O(T\cdot \log n)$. Step 1 also involves distributing the clusters to the correct database node, which is often an expensive I/O bound process. So we do spatial partitioning only once based on the bottom level, and reuse the same partition scheme for other levels to avoid moving data around database nodes. 
Step 2 runs in $O(M\log M)$ because the single node algorithm is run in parallel across partitions. Step 3 takes $O(n\log T)$ because there are $\log T$ KD-tree levels in total, and we need to consider for each KD-tree level $n$ clusters in the worst case. However, Step 3 is expected to run very fast in practice because most clusters are out of the range in Figure \ref{fig:mergepartition}.

\vspace{.3em}
\subhead{Other partitioning strategies}. One could partition the data using fields other than $x$ and $y$ and then in a similar fashion, run the single-node algorithm on the resulting partitions in parallel. However, since the two spatial attributes are not involved in partitioning, objects in each partition would span the whole 2D space. So even though overlap and density constraints are satisfied within each partition, when merged together, they will very likely be violated unless extra spatial postprocessing are in place. We therefore choose to perform spatial partitioning throughout to guarantee \textbf{G3-a} and \textbf{G3-b}. 


\subsection{Data Fetching} \label{subsec:datafetching}

The data fetcher's job is to efficiently fetch data in the user's viewport (\textbf{G4-b}). We make use of multi-node spatial indexes, which can help fetch objects in a viewport-sized region with interactive response times.

\vspace{.3em}
\subhead{Creating multi-node spatial indexes}. Suppose the $j$-th ($1\leq j\leq T$) partition on zoom level $i$ is stored in the database table $t_{i,j}$, which has roughly $M$ clusters. We augment all such $t_{i,j}$ with a box-typed column \texttt{bbox}, which stores the bounding box of cluster marks. We then build a spatial index on column \texttt{bbox}, by issuing the following query:
\begin{lstlisting}[basicstyle=\ttfamily\normalsize, frame = none, escapeinside = `']
CREATE INDEX sp_idx ON `$t_{i,j}$' using gist(bbox);
\end{lstlisting}
where \texttt{gist} is the spatial index based on the generalized search tree\cite{postgresgist}.
In practice, these \texttt{CREATE INDEX} statements can be run in parallel by the multi-node database. 

\vspace{.3em}
\subhead{Fetching data from relevant partitions}. Given a user viewport $V$ on zoom level $i$, clusters from partition $t_{i,j}$ that are inside $V$ can be fetched by a query like the following:
\begin{lstlisting}[basicstyle=\ttfamily\normalsize, frame = none, escapeinside = `']
SELECT * FROM `$t_{i,j}$' WHERE bbox && `$V$';
\end{lstlisting}
where \texttt{\&\&} is the intersection operator. The spatial index on \texttt{bbox} ensures that this query runs fast. We traverse the KD-tree to find out partitions that intersect $V$, run the above query on these partitions and union the results. Note that for top zoom levels that are small in size, there can be too many partitions that intersect with the viewport, which can be harmful for data fetching performance because we need to wait for sequential network trips to many database nodes. Therefore, we merge all partitions on each of the top $L$ levels into one database table. $L$ is an empirically determined constant based on the relative size of the zoom levels to the viewport size. 

\section{Implementation} \label{sec:impl}


We implement \sys as an extension to Kyrix\cite{tao2019kyrix}, a general pan/zoom system we have built. This enables the developer to both rapidly author \visnames and reuse features of a general pan/zoom system in one integrated system. For example, Kyrix supports multiple coordinated views. Without switching tools, the developer can construct a multi-view visualization in which one or more views are \visnames authored with \sys. As another example, the developer can augment \visnames with the semantic zooming functionality provided by Kyrix, where the user can click on a visual mark and zoom into another \visname. 
Furthermore, Kyrix provides APIs for integrating a pan/zoom visualization into a web application, which are highly desired by the \visname developers we collaborate with. Examples include programmatic pan/zoom control, notifications of pan/zoom events, getting current visible data items.

\vspace{.3em}
\subhead{Specification compilation}. \sys uses a \textit{Node.js} module to validate the JSON-based \visname specification. Validated specifications are compiled into low-level Kyrix specifications so that part of Kyrix's frontend code can be reused to handle rendering and pan/zoom interactions. 

\vspace{.3em}
\subhead{Layout generator and data fetcher}. \sys's layout generator and data fetcher override respectively Kyrix's index generator and data fetcher. Both components are written in the same Java application, using the Java Database Connectivity (JDBC) to talk to Citus\footnote{https://www.citusdata.com/}, an open-source multi-node database built on top of PostgreSQL. The layout generator uses PLV8\footnote{https://plv8.github.io/}, a PostgreSQL extension that enables implementation of algorithms in Section \ref{sec:algorithm} in Javascript functions, along with parallel execution of those functions directly inside each Citus database node.

\vspace{.3em}
\subhead{Database deployment and orchestration}. \sys provides useful scripts for one-command deployment of \sys and database dependencies (\textbf{G1}). We use Kubernetes\footnote{https://cloud.google.com/kubernetes-engine/} to orchestrate a group of nodes running containerized Citus and \sys built with Docker\footnote{https://www.docker.com/}. 

\begin{table*}[t]
\centering
\caption{Online serving time (95-th percentile, in milliseconds).}
\label{tab:serving}
\vspace{-.25em}
\resizebox{0.75\textwidth}{!}{%
\begin{tabular}{|c|c|c|c|c|c|}
\hline
 & \begin{tabular}[c]{@{}c@{}}\textsc{Reddit Text}\\ (Figure \ref{fig:gallery}e, \\ 1B objects)\end{tabular} & \begin{tabular}[c]{@{}c@{}}\textsc{Reddit Circle} \\ (Figure \ref{fig:teaser}, \\ 1B objects)\end{tabular} & \begin{tabular}[c]{@{}c@{}}\textsc{Taxi Heatmap}\\ (Figure \ref{fig:gallery}a, \\ 178.3M objects)\end{tabular} & \begin{tabular}[c]{@{}c@{}}\textsc{Taxi Contour}\\ (Figure \ref{fig:gallery}b, \\ 178.3M objects)\end{tabular} & \begin{tabular}[c]{@{}c@{}}\textsc{Liquor}\\ (Figure \ref{fig:gallery}d, \\ 17.3M objects)\end{tabular} \\ \hline
Data Fetching & 14 & 17 & 32 & 32 & 14 \\ \hline
Network & 1 & 1 & 223 & 254 & 1 \\ \hline
\end{tabular}%
}
\vspace{-.25em}
\end{table*}

\begin{table*}[t]
\centering
\caption{Offline indexing time (in minutes). }
\vspace{-.25em}
\label{tab:indexing}
\resizebox{0.85\textwidth}{!}{%
\begin{tabular}{|c|c|c|c|c|c|}
\hline
 & \begin{tabular}[c]{@{}c@{}}\textsc{Reddit Text}\\ (Figure \ref{fig:gallery}e, \\ 1B objects)\end{tabular} & \begin{tabular}[c]{@{}c@{}}\textsc{Reddit Circle} \\ (Figure \ref{fig:teaser}, \\ 1B objects)\end{tabular} & \begin{tabular}[c]{@{}c@{}}\textsc{Taxi Heatmap}\\ (Figure \ref{fig:gallery}a, \\ 178.3M objects)\end{tabular} & \begin{tabular}[c]{@{}c@{}}\textsc{Taxi Contour}\\ (Figure \ref{fig:gallery}b, \\ 178.3M objects)\end{tabular} & \begin{tabular}[c]{@{}c@{}}\textsc{Liquor}\\ (Figure \ref{fig:gallery}d, \\17.3M objects)\end{tabular} \\ \hline
Building KD-tree (Step 1) & 11.8 & 10.5 & 2.7 & 2.4 & 0.7 \\ \hline
Redistributing data (Step 1)& 94.3 & 100.0 & 8.5 & 8.4 & 1.3 \\ \hline
Parallel clustering (Step 2) & 9.9 & 3.7 & 6.9 & 9.0 & 4.7 \\ \hline
Merge partitions (Step 3) & 61.3 & 18.2 & 1.1 & 0.8 & 0.1 \\ \hline
Creating Spatial Indexes& 2.4 & 1.3 & 1.2 & 1.2 & 1.3 \\ \hline
Total & 179.7 & 133.8 & 20.3 & 21.8 & 8.2 \\ \hline
\end{tabular}%
}
\vspace{-1em}
\end{table*}
\section{Evaluation} \label{sec:evaluation}
\vspace{-.25em}
We conducted extensive experiments to evaluate two aspects of \sys: 1) performance and 2) authoring effort.

\subsection{Performance}
\vspace{-.25em}
We conducted performance experiments to evaluate the online serving and indexing performance of \sys. We used both example \visnames in Figures \ref{fig:teaser} and \ref{fig:gallery} and a synthetic circle-based \visname \textsc{Syn} that visualizes a skewed dataset where 80\% of the objects are in 20\% of the 2D plane, and the rest of the 20\% are uniformly distributed across the 2D plane. For database partitioning, we set $M=2$ million, i.e., each partition has roughly 2 million objects. So for a dataset with $N$ objects, there are $K=\left\lceil{\frac{N}{M}}\right\rceil$ partitions. Based on the number of partitions, we provision a Google Cloud Kubernetes cluster with $\left\lceil{\frac{K}{8}}\right\rceil$ \texttt{n1-standard-8} PostgreSQL nodes (8 vCPUs, 30GB memory), each serving 8 partitions. 

\subsubsection{Online Serving Performance}
\vspace{-.25em}
To measure the online response times, we used a user trace where one pans around to find the most skewed region on a zoom level, zooms in, repeats until reaching the bottom level and then zooms all the way back to the top level. We measured the 95-th percentile\footnote{A 95-percentile says that 95\% of the time, the response time is equal to or below this value. This is a common metric for measuring network latency of web applications. } of all data fetching time and network time. 


Table \ref{tab:serving} shows the results on five \visnames. The 95-percentile data fetching times were all below 32ms. The reason was because we only fetched data from the partitions that intersect with the viewport and the spatial indexes sped up the spatial queries. Network times were mostly negligible except for \textsc{Taxi Heatmap} and \textsc{Taxi Contour}, where many more data items were fetched due to smaller $\theta$ values. 

\begin{figure}[t]
\centering
\vspace{-.25em}
\includegraphics[width=0.4\textwidth]{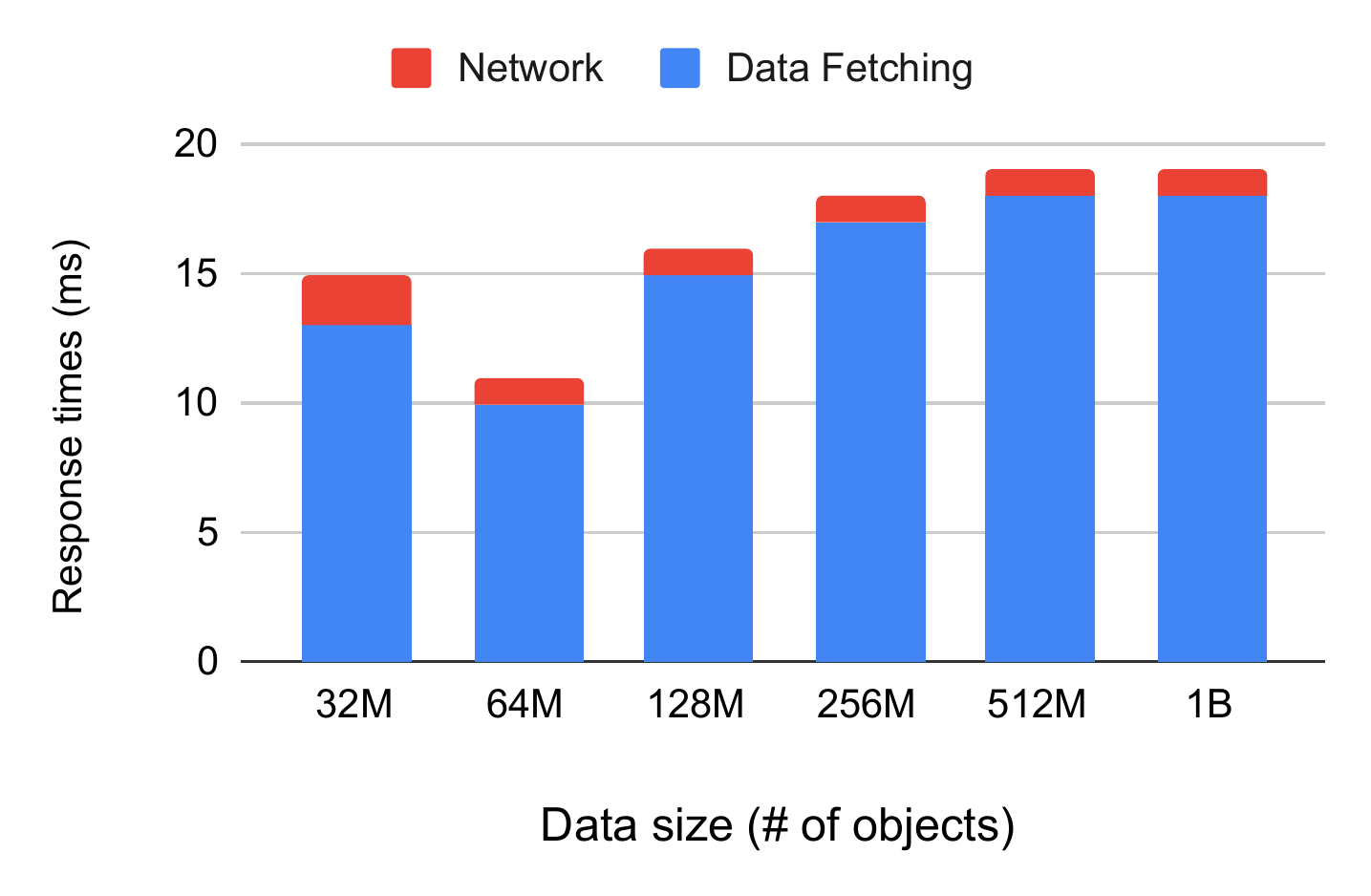}
\vspace{-1.5em}
\caption{Serving scalability on the synthetic \visname \textsc{Syn}.\label{fig:servingscalability}}
\vspace{-1em}
\end{figure}

Figure \ref{fig:servingscalability} shows the response times on different sizes of the synthetic \visname \textsc{Syn}. We can see that the response times remained stably under 20ms for data sizes from 32 million to 1 billion. 

\subsubsection{Offline Indexing Performance}

\begin{figure}[t]
\centering
\includegraphics[width=0.45\textwidth]{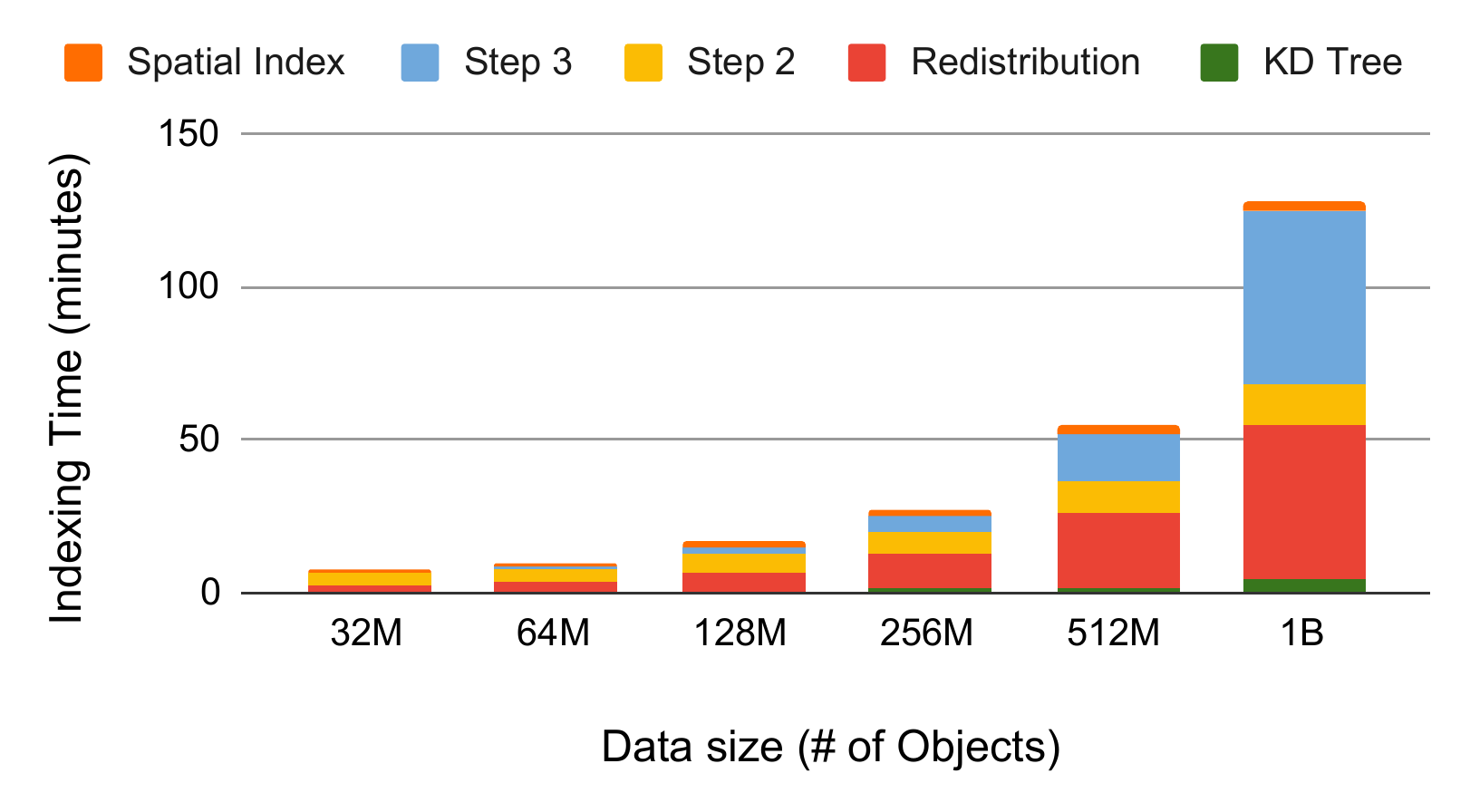}
\vspace{-1.5em}
\caption{Indexing scalability on the synthetic \visname \textsc{Syn}.\label{fig:indexingscalability}}
\vspace{-1.5em}
\end{figure}

Table \ref{tab:indexing} shows the indexing performance of the layout generator on five example \visnames. We make the following observations. First, the indexing phase finished in reasonable time: every example finished in less than 3 hours. Second, redistributing the data to the correct spatial partition was the most time consuming part since it was an I/O bound process. Fortunately, the same spatial partitions can be reused for updatable data if the spatial distribution does not change drastically. Third, parallel clustering and spatial index creation took the least time because they could be run in parallel across partitions. Fourth, merging clusters along KD-tree splits was mostly a cheap process. In fact, the largest number of clusters along a KD-tree split was 16,647. The reason that this step took longer on \textsc{Reddit Text} than on \textsc{Reddit Circle} was because it had more zoom levels (20 vs. 15) due to larger mark size (text vs. circle). Moreover, iterating through objects along KD-tree splits were much more time-consuming on the bottom five levels. 

Figure \ref{fig:indexingscalability} shows how indexing time changed for different sizes of \textsc{Syn}. We can see that the indexing time scaled well as the data size grew: as data size doubled, indexing time roughly doubled as well.

\subsection{Authoring Effort}\label{subsec:authoringexp}
To evaluate the accessibility of our grammar, we compared the authoring effort of \sys with Kyrix\cite{tao2019kyrix}, a state-of-the-art general pan/zoom system. To our best knowledge, Kyrix is the only system that offers declarative primitives for general pan/zoom visualizations, and has been shown to be accessible to visualization developers in a user study\cite{tao2019kyrix}. Former systems/languages such as D3\cite{d3}, Pad++\cite{bederson1994pad++}, Jazz\cite{bederson2003jazz} and ZVTM\cite{pietriga2005toolkit} require procedural programming which generally takes more authoring effort\cite{tao2019kyrix}. We measured lines of specifications using both systems for the two examples \visnames in Figures \ref{fig:gallery}d and \ref{fig:gallery}f. We used a code formatter\footnote{https://prettier.io/} to standardize the specifications, and only counted non-blank and non-comment lines.\footnote{Code in this experiment is included in the supplemental materials.}

\begin{table}[h]
\centering
\vspace{-.25em}
\caption{Comparison of lines of specifications when using \sys and Kyrix to author the two example \visnames in Figure \ref{fig:gallery}d and Figure \ref{fig:gallery}f. \label{tab:authoring}}

\resizebox{0.45\textwidth}{!}{%
\renewcommand{\arraystretch}{1.2}
\begin{tabular}{|c|c|c|c|}
\hline
 & \sys & Kyrix & \begin{tabular}[c]{@{}c@{}}\sys's saving\\ over Kyrix\end{tabular} \\ \hline
Figure \ref{fig:gallery}d & 62 lines & 568 lines & $9.2\times$ \\ \hline
\begin{tabular}[c]{@{}c@{}}Figure \ref{fig:gallery}f\\ w/ custom renderer\end{tabular} & 164 lines & 610 lines & $3.7\times$ \\ \hline
\begin{tabular}[c]{@{}c@{}}Figure \ref{fig:gallery}f\\ w/o custom renderer\end{tabular} & 68 lines & 514 lines & $7.6\times$ \\ \hline
\end{tabular}%
}
\vspace{-.75em}
\end{table}

Table \ref{tab:authoring} shows the results. We can see that when authoring the two example \visnames, \sys achieved respectively $9.2\times$ and $3.7\times$ saving in specifications compared to Kyrix. In the second example, when we excluded the custom renderer for soccer players (which has 96 lines), the amount of savings was $7.6\times$. These savings came from \sys abstracting away low-level details such as rendering of visual marks, configuring zoom levels, etc. 

The above comparison did not include the code for layout generation. To enable the comparison, we stored the layouts generated by \sys as database tables so that Kyrix could directly use them. However, programming the layout was in fact a challenging task, as indicated by the total lines of code of \sys's layout generator (1,439). Therefore, we conclude that \sys greatly reduced the user's effort in authoring \visnames compared to general pan/zoom systems. 

\section{Limitations and Future Work} \label{sec:discussion}
\subhead{Other layout strategies}. \sys's assumes that the location of a mark comes from an object. This can be relaxed to diversify our layout generator. For example, supporting inexact placement of marks such as binned aggregation\cite{heimerl2018visual} in \visnames is one future direction. We also plan to investigate layout strategies that concern multi-class scatterplots, e.g. how to preserve relative density orders among multiple classes\cite{chen2014visual,chen2019recursive}. 

\vspace{.3em}
\subhead{More built-in templates}. Our declarative grammar is designed to enable rapid extension of the system with custom marks. This motivates us to engage more with the open-source community and enrich our built-in mark gallery with  templates commonly required/authored by developers. 

\vspace{.3em}
\subhead{Incremental updates}. Currently, \sys assumes that data is static and pre-materialize mark layouts. To interactively debug, the developer needs to either use a sample of the data or reduce the number of zoom levels. It is our future work to identify ways to incrementally update our mark layout upon frequent changes of developer specifications, as well as when the data itself is updated dynamically. 

\vspace{.3em}
\subhead{Animated transitions}. A discrete-zoom-level model simplifies layout generation, but can potentially lead to abrupt visual effect upon level switching, especially for KDE-based renderers such as heatmaps. As future work, we will use animated transitions to counter this limitation. 

\vspace{.3em}
\subhead{Raster Images-based \visnames}. The visual density constraint, partly due to limited processing capabilities of the frontend and the database, forbids the creation of dense visualizations such as point clouds\cite{cartolabe}. We envision the use of raster images to remove this constraint for these visualizations where interaction with objects is not required.


\section{Conclusion} \label{sec:conclusion}

In this paper, we presented the design of \sys, a system for easy authoring of \visnames at scale. \sys contributed a declarative grammar that enabled concise specification of a wide range of \visnames and rapid authoring of custom marks. Behind the scenes, \sys automatically generated layout of visual marks on zoom levels using a range of usability guidelines such as maintaining a visual density budget and high data abstraction quality. To scale to big datasets, \sys worked with a multi-node parallel database system to implement the layout algorithm in a distributed setting. Multi-node spatial indexes were built to achieve interactive response times. We demonstrated the expressivity of \sys with a gallery of example \visnames. Experiments on real and synthetic datasets showed that \sys scaled to big datasets with billions of objects and reduced the authoring effort significantly compared to a state-of-the-art authoring system. 


\vspace{-.5em}
\section{Acknowledgement}
\vspace{-.5em}
We thank the anonymous reviewers for their thoughtful feedback. This work was in part supported by NSF OAC-1940175, OAC-1939945,  IIS-1452977, DGE-1855886, IIS-1850115,
DARPA FA8750-17-2-0107 and the Data Systems and AI Lab initiative under Grant 3882825.



\bibliographystyle{abbrv-doi}

\bibliography{bib/general_pan_zoom, bib/zsv, bib/static_sv, bib/grammar, bib/misc}

\clearpage


\end{document}